\documentclass[12pt,english,twoside]{article}
\usepackage[dvips]{graphicx}
\usepackage{authblk}
\usepackage{subeqnarray}
\usepackage{dsfont}
\usepackage{caption}
\usepackage{subcaption}
\usepackage{lipsum}
\usepackage{array,multirow,makecell}
\setcellgapes{1pt}
\RequirePackage{color,ragged2e,newlfont,pifont,stmaryrd,xspace}
\RequirePackage{amsmath,amssymb,amsfonts,textcomp,yhmath}
\RequirePackage{times} 
\RequirePackage{cases} 
\usepackage[]{hyperref}
\hypersetup{backref,implicit=true,bookmarks=true, colorlinks=true,
linkcolor=blue,filecolor=blue,
pagecolor=blue,urlcolor=blue,pdfpagemode=UseOutlines}

\newtheorem{theorem}{Theorem}[section]

\newtheorem{proposition}[theorem]{Proposition}

\newtheorem{remark}{Remark}

\newenvironment{prev}[1][\underline{Proof}]{\textbf{#1.} }{\ \rule{0.5em}{0.5em}}
\begin{document}
\markboth{\small{A. Nangue}}{\tiny{Global stability analysis of the
original cellular model of hepatitis C Virus infection under
therapy}}

\title{Global stability analysis of the
original cellular model of hepatitis C Virus infection under
therapy}
\date{}
\author[,1]{Alexis Nangue \thanks{Email : Alexis Nangue : \texttt{alexnanga02@yahoo.fr}}}

\affil[1]{University of Maroua, Higher Teacher's Training College,
Department of Mathematics, P.O.Box 55 Maroua, Cameroon}

\maketitle

\begin{abstract}
In this paper, we present the global analysis of a HCV model under
therapy. We prove that the solutions with positive initial values
are global,  positive, bounded and not display periodic orbits. In
addition, we show that the model is globally asymptotically stable,
by using appropriate Lyapunov functions.
\end{abstract}
{\bf Keywords}:HCV cellular model, differential system, therapy,
local and global solution, invariant set, stability.

\section{Introduction}
According to \cite{who} recent estimates, more than 185 million
people around the world have been infected with the hepatitis C
virus (HCV), of whom 350 000 die each year. One third of those who
become chronically infected are predicted to develop liver cirrhosis
or hepatocellular carcinoma. Despite the high prevalence of disease,
most people infected with the virus are unaware of their infection.
For many who have been diagnosed, treatment remains unavailable.
Treatment is successful in the majority of persons treated, and
treatment success rates among patients treated in low- and
middle-income countries are similar to those in high-income
countries Hepatitis C virus (HCV) infects liver cells (hepatocytes).
Approximately 200 million people worldwide are persistently infected
with the HCV and are at risk of developing chronic liver disease,
cirrhosis and hepatocellular carcinoma. HCV infection therefore
represents a significants global public health problem. HCV
established chronic hepatitis in $60\%$-$80\% $ of infected adults
\cite{seeff}.\\
\indent In literature, several mathematical models have been
introduced for understanding HCV temporal dynamics \cite{dahari,
neuman,
reluga}.\\
\indent In this article, we consider the basic extracellular model
with therapy presented by Neumann et al. in \cite{neuman}. Given the
recent surge in the development of new direct acting antivirals
agents for HCV therapy, mathematical modelling of viral kinetics
under treatment continues to play an instrumental role in improving
our knowledge and understanding of virus pathogenesis and in guiding
drug development \cite{chartej, guedj, Rong}.
\\\indent
To proceed, we assume that the uninfected target cells are produced
at a rate $\lambda$, die at constant rate $d$ per cell. On the other
hand, the target cells are infected with de novo infection rate
constant of $\beta$ and the infected cells die at a constant rate of
$\delta$ per cell. The hepatitis C virions are produced inside the
infected cells at an average rate $p$ per infected cell and have a
constant clearance rate $c$ per virion. Thereby, viral persistence
will occur when rate of viral production $(p)$, de novo infection
$(\beta)$, and production of target cells $(\lambda)$ exceeds the
clearance rate $(c)$, death rate of infected cells $(\delta)$ and
target cells death rate $(d)$. In addition, the therapeutic effect
of IFN treatment in this model involved blocking virions production
 and reducing new infections which, are described in fractions
 $(1-\varepsilon)$ and $(1-\eta)$, respectively.
 ($0\leq\varepsilon \leq 1$,
 $ 0\leq \eta \leq 1$).
\begin{figure}[!h]
\centering
\includegraphics[angle=0,height=5cm,width=10cm]{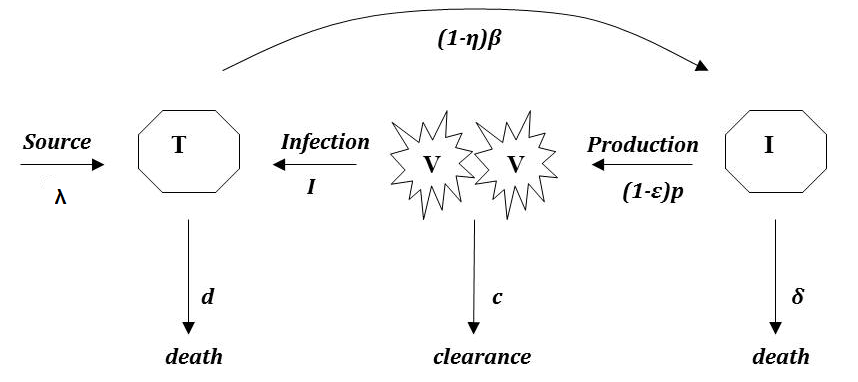}
\caption{Schematic representation of the original viral kinetic
model of HCV infection under treatment. T represents the target
uninfected cells, I is the infected cells and V represents the free
virus }
\end{figure}
According to \cite{chong, neuman}, the above assumptions lead to the
following differential equations :
\begin{subequations}\label{dyna:gp}
\begin{align}
\frac{d T}{dt}&= \lambda-dT-{\color{red}(1-\eta)}\beta V T; \label{dyna:gp1} \\
  \frac{d I}{dt}&={\color{red}(1-\eta)}\beta V T-\delta I;  \label{dyna:gp2} \\
  \frac{d V}{dt}&={\color{red}(1-\varepsilon)}pI-cV ;\label{dyna:gp3}
\end{align}
\end{subequations}
 where the equations relate the dynamics relationship between, T as
the uninfected target cells (hepatocytes), I as the infected cells
and V as the viral load (amount of viruses present in the blood). In
this article, model system (\ref{dyna:gp}) is taken as the original
model used to analyse the HCV dynamics.
\\\indent
The initial conditions associated to system (\ref{dyna:gp}) are
given by :
\begin{equation}\label{condinit}
 T(t_{0})=T_{0},\; I(t_{0})=I_{0}, \; V(t_{0})=V_{0} ,\;\; t_{0}\in
[0, +\infty[  .
\end{equation}
This paper is organized as follows : The global properties of the
solutions to the mathematical model is carried out in Section 1. The
stability of the disease non-infected steady state, and the infected
steady state is analysed in section 2.
\section{Properties of solutions to the Cauchy problem (\ref{dyna:gp}),(\ref{condinit}) }
\subsection{Existence of local solutions}
The first step in examining model (\ref{dyna:gp}) is to prove that
local solution to the initial-value problem does, in fact, exist,
and that this solution is unique.
\begin{theorem}
Let $T_{0}$, $I_{0}$, $V_{0}$ $ \in \mathds{R} $ be given. There
exists $t_{1} > t_{0} > 0$ and continuously differentiable functions
$T$, $I$, $V$ $: [0, t_{0}) \longrightarrow \mathds{R}$ such that
the ordered triple $(T, I, V )$ satisfies (\ref{dyna:gp}) and
$(T(t_{0}), I(t_{0}), V(t_{0}))=(T_{0}, I_{0}, V_{0})$.
\end{theorem}
\begin{prev}To prove the result, we use the classical
Cauchy-Lipschitz theorem. Since the first order system of ordinary
differential equations (\ref{dyna:gp}) is autonomous, it suffices to
show that the function $H : \mathds{R}^{3} \longrightarrow
\mathds{R}^{3}$ defined by :
\begin{equation}\label{formsimple}
H(x)=\left(
       \begin{array}{c}
         h_{1}(x_{1}, x_{2}, x_{3}) \\
         h_{2}(x_{1}, x_{2}, x_{3}) \\
         h_{3}(x_{1}, x_{2}, x_{3}) \\
       \end{array}
     \right)
     =
     \left(
       \begin{array}{c}
         \lambda-d x_{1}-(1-\eta)\beta x_{3}x_{1} \\
          (1-\eta)\beta x_{3}x_{1}-\delta x_{2}\\
          (1-\varepsilon)p x_{2}-c x_{3}\\
       \end{array}
     \right)
\end{equation}
is locally Lipschitz in its $x$ argument. In fact, it is enough to
notice that the jacobian matrix
\begin{equation*}
    \nabla H(x)= \left[
                   \begin{array}{ccc}
                  -d -(1-\eta)\beta x_{3}    & 0 & -(1-\eta)\beta x_{1} \\
                   (1-\eta)\beta x_{3}   & -\delta & (1-\eta)\beta x_{1} \\
                   (1-\varepsilon)p   & 0 & -c \\
                   \end{array}
                 \right]
\end{equation*}
is linear in $x$ and therefore locally bounded for every $ x \in
\mathds{R}^{3}$. Hence, H has a continuous, bounded derivative on
any compact subset of $\mathds{R}^{3}$ and so H is locally Lipschitz
in $x$; In addition H is continuous. By Cauchy-Lipschitz theorem,
there exits a unique solution, $x(t)$, to the ordinary differential
equation $$x'(t)=H(x(t))$$ with initial value $x(t_{0})=x_{0}$ on
$[t_{0}, t_{1}]$ for some time $t_{1}>t_{0}\geq 0$.
\end{prev}
\begin{remark}
The model (\ref{dyna:gp}) can be rewrite in the form
\begin{equation*}
    S'(t)=H(S(t))
\end{equation*}
where $S(t)=(T(t), I(t), V(t))^{t}$ and F is defined by
(\ref{formsimple}).
\end{remark}
\begin{remark}
Since H is a continuously differentiable function, we  deduce a
unique maximal solution of initial value problem (\ref{dyna:gp}),
(\ref{condinit}). In addition, F, being indefinitely continuously
differentiable, we can also deduce that this solution is also only
if indefinitely continuously differentiable.
\end{remark}
Additionally, we may show that for positive initial data, solutions
of (\ref{dyna:gp}), (\ref{condinit}) remain positive as long as they
exist.
\subsection{Positivity}
\begin{theorem}\label{theopositivity}
Let $(T, I, V)$ be a solution of the Cauchy problem (\ref{dyna:gp}),
(\ref{condinit}) on an interval $[t_{0}, t_{1}[$. Assume the initial
data of (\ref{dyna:gp}), (\ref{condinit}) satisfy $T_{0} > 0$,
 $I_{0} >
0$, and $V_{0} > 0$ then $T(t)$, $T(t)$ and $V(t)$ remain positive
for all $t \in [t_{0}, t_{1}[$.
\end{theorem}
\begin{prev}
Call the variables $x_{i}$. If there is an index $i$ and a time $t$
 for which $x_{i}(t) = 0$,  let $t_{*}$ be the infimum of all such
  $t$ for any i.
Then the restriction of the solution to the interval $[t_{0};
t_{*}[$ is positive and $x_{i}(t_{*}) = 0$ for a certain value
 of $i$.
The equation for $x_{i}$ in the system (\ref{dyna:gp}) can be
written in the form :
 $$\frac{dx_{i}(t)}{dt}= -x_{i}f(x) + g(x),$$ where $g(x)$ is non-negative. As a
  consequence $\frac{d x_{i}(t)}{dt}\geq -x_{i}f(x) $ and
  $x_{i}(t) > 0$, $ \forall t \in [t_{0}, t_{*}] $.
 In fact :\\
Recall that all constants in the system (\ref{dyna:gp}) are
non-negative. Using this and the solutions on $[t_{0}, t^{*} [$, we
have :
\begin{equation*}
    \left\{
      \begin{array}{ll}
 \frac{d I}{dt}=-I \delta +(1-\eta)\beta V T& \hbox{} \\
 \frac{d V}{dt}=-Vc+(1-\varepsilon)pI& \hbox{}
      \end{array}
    \right.
\Rightarrow \left\{
      \begin{array}{ll}
 \frac{d I}{dt} \geq -I \delta & \hbox{} \\
 \frac{d V}{dt}\geq - V c& \hbox{}
      \end{array}
    \right.
\Rightarrow \left\{
      \begin{array}{ll}
 \frac{d}{dt}(\ln I(t)) \geq - \delta & \hbox{} \\
 \frac{d}{dt}(\ln V(t))\geq -  c ,& \hbox{}
      \end{array}
    \right.
\end{equation*}
which yields : \begin{eqnarray*}
   I(t)  &\geq & I_{0}e^{-\delta(t-t_{0})} , \\
    V(t) &\geq & V_{0}e^{-c(t-t_{0})} \;\; \mbox{for all $t\in [t_{0}, t^{*}]$ }.
\end{eqnarray*}
Similarly, in one hand we have :
\begin{equation*}
\frac{d T}{dt} = \lambda-dT-{\color{red}(1-\eta)}\beta V T \leq
\lambda.
\end{equation*}
Solving for T yields
\begin{equation*}
 T(t) \leq  T_{0}+\lambda(t-t_{0}) \leq  A_{1}(1+(t-t_{0}))
\end{equation*}
where $A_{1} \geq \max\{T_{0}, \lambda\} $. In other hand we have :
\begin{equation*}
    \frac{d}{dt}(I+V)=-I \delta +(1-\eta)\beta V T-Vc+(1-\varepsilon)pI
    \leq \beta V T+ p I.
\end{equation*}
Recall that we have a bound on T, so\begin{eqnarray*}
 \frac{d}{dt}(I+V)  &\leq& A_{1}(1+(t-t_{0}))\beta T +p I \\
                    &\leq& A_{2}(1+(t-t_{0}))( T +I)
\end{eqnarray*}
where $ A_{2}\geq \max\{A_{1}\beta, p\} $. Solving the differential
inequation yields :
\begin{equation}\label{h1t}
    I(t)+V(t) \leq A_{3}H_{1}(t), \;\; \forall t \in [t_{0}, t^{*}]
\end{equation}
where $A_{3} > 0 $ depends upon $A_{2}$, $I_{0}$ and $V_{0}$ only,
and $H_{1}(t)=e^{\int_{t_{0}}^{t}(1+(t-t_{0}))d\tau} > 0 $, $t \in
[t_{0}, t^{*}]$. Using the fact that $I(t)$ and $V(t)$ are positive,
(\ref{h1t}) yields :
\begin{eqnarray*}
  A_{3}H_{1}(t)\geq I(t)+V(t) &\geq & V(t) , \\
  A_{3}H_{1}(t) \geq I(t)+V(t) & \geq&  I(t).
\end{eqnarray*}
With these bounds in place, we can now examine $T(t)$ and bound it
from below using :
\begin{eqnarray*}
   \frac{d T}{dt}&=& \lambda-dT-(1-\eta)\beta V T \\
   &\geq & -dT-(1-\eta)\beta V T \\
   & \geq & -dT-(1-\eta)\beta A_{3}H_{1}(t) T \\
\frac{d T}{dt} & \geq& -A_{4}(1+H_{1}(t))T
\end{eqnarray*}
for $t \in [t_{0}, t^{*}] $, where $A_{4} \geq \max\{d,
(1-\eta)\beta A_{3}\} $. Shifting that last term to the other side
of the equation yields :
\begin{equation*}
    \frac{d T(t)}{dt}+A_{4}(1+H_{1}(t))T(t) \geq 0.
\end{equation*}
Since we know
\begin{equation*}
    \dfrac{d}{dt}\left( T(t)+e^{A_{4}\int_{t_{0}}^{t}(1+H_{1}(\tau))d\tau}\right)\geq 0,
\end{equation*}
we find that for $t \in [t_{0}, t^{*}] $,
\begin{equation*}
    T(t)\geq e^{-A_{4}\int_{t_{0}}^{t}(1+H_{1}(\tau))d\tau}> 0
\end{equation*}
Therefore $x_{i}(t)> 0$, for all $t \in [t_{0}, t^{*}] $. In
particular  $x_{i}(t^{*})> 0$, which is a contradiction and the
theorem is proven.
\end{prev}
\begin{remark}
\begin{enumerate}
    \item[(i)]With this, we have a general idea that the model is sound,
and can say with certainty that it remains biologically valid as
long as it began with biologically-reasonable (i.e, positive) data.
This also shows that once infected, it is entirely possible that
 the virus may
continue to exist beneath a detectable threshold without doing any
damage.
    \item[(ii)]
One reason why we choose the strict inequalities for the initial
data is that often in biological (or chemical) applications we are
interested in the case of solutions where all unknowns are positive.
This means intuitively that all elements of the model are 'active'.
On the other hand it is sometimes relevant to consider solutions
with non-strict inequalities. In fact the statement of the
theorem\label{theopositivity} with strict inequalities implies the
corresponding statement with non-strict inequalities in a rather
easy way.
\end{enumerate}
\end{remark}
\subsection{Existence of global solutions}
It will now be shown, with the help of the continuation criterion
the existence of global solutions.
\begin{theorem}
The solutions  of the Cauchy problem (\ref{dyna:gp}),
(\ref{condinit}), with positive initial data,
 exist globally in time
in the future that is : on $[t_{0}, +\infty[$.
\end{theorem}
\begin{prev}
To prove this it is enough to show that all variables are bounded on
an arbitrary finite interval $[t_{0}; t)$. Using the positivity of
the solutions is suffices to show that all variables are bounded
above.\\\indent
 Taking the sum
of  equations (\ref{dyna:gp1}) and (\ref{dyna:gp2}) shows that :
$$ \frac{d}{dt}(T+I) \leq \lambda $$
 and hence that $T(t)+I(t)\leq  T_{0}+I_{0}+\lambda(t-t_{0})$.
 Thus $T$ and $I$ are
bounded on any finite interval. The third equation i.e. equation
(\ref{dyna:gp3}), then shows that $V(t)$ cannot grow faster than
linearly and is also bounded
 on any finite interval.
\end{prev}
\subsection{Global boundedness of solutions}
\begin{theorem}\label{globound}
For any positive solution $(T, I, V)$ of system
(\ref{dyna:gp}), (\ref{condinit})  we have :
\begin{equation*}
    T(t)+I(t) \leq C_{1} \; and \; V(t)\leq C_{2}
\end{equation*}
where
\begin{equation*}
    C_{1}=\max\left\{T_{0}+I_{0}, \frac{\lambda}{\min\{d, \delta\}}\right\},
    C_{2}=\max\left\{V_{0}, \frac{(1-\varepsilon)pC_{1}}{c}\right\}.
\end{equation*}
\end{theorem}
\begin{prev}
According to equations (\ref{dyna:gp1}) and (\ref{dyna:gp2}), we
have :
\begin{eqnarray*}
  \frac{d}{dt}(T+I) &=& \lambda -dT-\delta I \\
   &\leq & \lambda -\min\{d,\delta\}(T+I).
\end{eqnarray*}
Gronwall inequality\cite{gronwall} yields :
\begin{eqnarray*}
T+I &\leq & (T(t_{0})+I(t_{0}))e^{-\min\{d, \delta\}
(t-t_{0})}+\int_{t_{0}}^{t}\lambda e^{-\int_{u}^{t}
-\min\{d, \delta\}dr}du  \\
   &\leq & (T_{0}+I_{0})e^{-\min\{d, \delta\}
(t-t_{0})}+\lambda\int_{t_{0}}^{t}e^{
-\min\{d, \delta\}(t-u)}du  \\
   & \leq & (T_{0}+I_{0})e^{-\min\{d, \delta\}
(t-t_{0})}+\lambda \dfrac{e^{-\min\{d, \delta\} (t-t)}
-e^{-\min\{d,\delta\} (t-t_{0})}}{\min\{d, \delta\}}
\end{eqnarray*}
\begin{eqnarray*}
   & \leq& \max\left\{(T_{0}+I_{0}), \frac{\lambda}{\min\{d, \delta\}}\right\}
   e^{-\min\{d, \delta\}
(t-t_{0})} \\
    & & +\max\left\{(T_{0}+I_{0}),\frac{\lambda}{\min{d,
\delta}}\right\}(1 -e^{-\min\{d,\delta\}
(t-t_{0})})  \\
   &\leq & \max\left\{(T_{0}+I_{0}),\frac{\lambda}
   {\min\{d,\delta\}}\right\}\left(
   e^{-\min\{d, \delta\}
(t-t_{0})} +1 - e^{-\min\{d, \delta\}
(t-t_{0})}\right) \\
T+I & \leq& \max\left\{(T_{0}+I_{0}),\frac{\lambda}
   {\min\{d,\delta\}}\right\}.
\end{eqnarray*}
Another hand, from equation (\ref{dyna:gp3}), we have :
\begin{eqnarray*}
  \frac{d V}{dt} &=& (1-\varepsilon)p I - c V  \\
   &\leq& (1-\varepsilon)p (I+T) - c V \\
   &\leq& (1-\varepsilon)p \max\left\{(T_{0}+I_{0}),\frac{\lambda}
   {\min\{d,\delta\}}\right\}  - c V  \\
   &\leq& (1-\varepsilon)p C_{1} - c V.
\end{eqnarray*}
Once more Gronwall inequality yields :
\begin{eqnarray*}
  V(t) &\leq& V(t_{0})e^{-c(t-t_{0})}+\int_{t_{0}}^{t}
  (1-\varepsilon)pC_{1}e^{\int_{u}^{t}cdr}du  \\
   &\leq& V_{0}e^{-c(t-t_{0})}+(1-\varepsilon)pC_{1}
   \int_{t_{0}}^{t}
  e^{-c(t-u)}du \\
   &\leq& V_{0}e^{-c(t-t_{0})}+(1-\varepsilon)pC_{1}
   \dfrac{e^{-c(t-t)}-e^{-c(t-t_{0})}}{c}  \\
   &\leq& V_{0}e^{-c(t-t_{0})}+(1-\varepsilon)pC_{1}
   \dfrac{1-e^{-c(t-t_{0})}}{c} \\
   &\leq& \max \left\{V_{0}, \frac{(1-\varepsilon)pC_{1}}{c}\right\}
   \left(e^{-c(t-t_{0})}+1-e^{-c(t-t_{0})}\right)  \\
  V(t) &\leq& \max \left\{V_{0}, \frac{(1-\varepsilon)pC_{1}}{c}\right\}
\end{eqnarray*}
Therefore the theorem is proven.
\end{prev}\\
As consequences of Theorem \ref{globound} we have :
\begin{remark}
Let S be a solution of system (\ref{dyna:gp}). If $S_{0} \in
\mathds{R}\times \mathds{R}^{3}_{+} $ then, the limit of $S(t)$
exits when $t \longrightarrow +\infty$ . In other words the solution
is globally uniformly bounded in the future. In particular, S is
periodic if and only if S is stationary under the condition that
$S(t)$ admits a finite limit when $t$ tends to infinity.
\end{remark}
\begin{theorem}Let $(t_{0}, S_{0}=(T_{0}, I_{0}, V_{0})) \in \mathds{R}\times \mathds{R}^{3}_{+}
$ and $([t_{0}, T[,  S=(T, I, V))$ be a maximal solution of the
Cauchy problem (\ref{dyna:gp}), (\ref{condinit})($T\in ]t_{0},
+\infty[$). If $T(t_{0})+I(t_{0})\leq C_{3}$ and  $V(t_{0})\leq
C_{4}$ then the set :
  $$ \Omega = \left\{(T(t), I(t), V(t))\in \mathds{R}^{3}_{+} :
  T(t)+I(t)\leq C_{3},\; V(t)\leq C_{4} \right\} ,$$
  where $C_{3}= \dfrac{\lambda}{\min\{d, \delta\}}$ and
  $C_{4}=\dfrac{(1-\varepsilon)p \lambda}{c \min\{d, \delta\}}$,
  is a positively invariant set by system (\ref{dyna:gp}).
\end{theorem}
\begin{prev}
Let $t_{1} \in [t_{0}, T[$. We shall show that :
\begin{enumerate}
    \item[(i)]If $T(t_{1})+I(t_{1})\leq C_{3}$ then for all $t_{1}\leq t < T$, $T(t)+I(t)\leq C_{3}$.
    \item[(ii)]If $V(t_{1})\leq C_{4}$ then for all $t_{1}\leq t < T$,  $V(t)\leq
    C_{4}$.
\end{enumerate}
\begin{enumerate}
    \item[(i)]Let us suppose that there exists
    $\varepsilon_{1} > 0$
    such that : $t_{1}\leq t_{1}+\varepsilon_{1} < T $,
     $$(T+I)(t_{1}+\varepsilon_{1}) >  C_{3} .$$
    Let $t^{*}_{1}=\inf\{t \geq t_{1}\, / \, (T+I)(t) >
     C_{3} \}$.\\
    Since $(T+I)(t^{*}_{1}) =  C_{3}$, hence\\
$(T+I)(t) =  C_{3} +
\frac{d}{dt}(T(t^{*}_{1})+I(t^{*}_{1}))(t-t^{*}_{1})+o(t-t^{*}_{1})$,
$t\longrightarrow t^{*}_{1}. $ In addition, according to equations
(\ref{dyna:gp1}) and (\ref{dyna:gp2}) of system (\ref{dyna:gp}) we
have :
$$ \frac{d}{dt}(T(t)+I(t))= \lambda -dT - \delta I $$
which
yields : \begin{eqnarray*}
 \frac{d}{dt}(T+I)(t^{*}_{1})  &\leq& \lambda - \min\{d, \delta\}(T+I)(t^{*}_{1})  \\
   &\leq& \lambda - \min\{d, \delta\}C_{3} \\
   &\leq& \lambda - \min\{d,\delta\}\dfrac{\lambda}{\min\{d, \delta\}} \\
   &\leq& 0,
\end{eqnarray*}
hence, there exists $\widetilde{\varepsilon}>0$ such that for all $
t^{*}_{1}\leq t < t^{*}_{1} + \widetilde{\varepsilon} $, $(T+I)(t)
\leq C_{3}$, a contradiction. therefore for all $t \in [t_{0}, T[$,
$(T+I)(t) \leq C_{3} $.
    \item[(ii)]
Let us suppose that there exists $\varepsilon_{1} > 0$
    such that : $t_{1}\leq t_{1}+\varepsilon_{1} < T $, $$V(t_{1}+\varepsilon_{1}) >  C_{4} .$$
    Let $t^{*}_{1}=\inf\{t \geq t_{1}\, / \, V(t) >  C_{3} \}$.\\
    Since $V(t^{*}_{1}) =  C_{3}$, hence\\
$V(t) =  C_{4} + \frac{d V(t^{*}_{1})
}{dt}(t-t^{*}_{1})+o(t-t^{*}_{1})$, $t\longrightarrow t^{*}_{1}. $
In addition, according to equation (\ref{dyna:gp3}) of system
(\ref{dyna:gp}) we have :
$$ \frac{d V(t)}{dt}= (1-\varepsilon)pI-cV $$
which yields : \begin{eqnarray*}
 \frac{d}{dt}V(t^{*}_{1})  &\leq& (1-\varepsilon)p(I+T)(t_{1}^{*})-cV(t_{1}^{*})  \\
   &\leq& (1-\varepsilon)pC_{3}-c C_{4} \\
   &\leq& (1-\varepsilon)p \frac{\lambda}{\min\{d,\delta\}}-c \frac{(1-\varepsilon)p \lambda}{c \min\{d, \delta\}} \\
\frac{d}{dt}V(t^{*}_{1})   &\leq& 0,
\end{eqnarray*}
hence, there exists $\widetilde{\varepsilon}>0$ such that for all $
t^{*}_{1}\leq t < t^{*}_{1} + \widetilde{\varepsilon} $, $V(t) \leq
C_{4}$, a contradiction. therefore for all $t \in [t_{0}, T[$, $V(t)
\leq C_{3} $.
\end{enumerate}
\end{prev}
\section{Stability analyses}
\subsection{Equilibria, Basic reproduction number $R_{0}$ and
local stability} According to \cite{chong}, apart from an
infection-free equilibrium
\begin{equation}\label{inffree}
E^{0}=(T^{0},0,0) \;\;where\; T^{0}=\frac{\lambda}{d}
\end{equation}
the system (\ref{dyna:gp}) has an infected equilibrium during
therapy $E^{*}=(T^{*}, I^{*}, V^{*})$, where :
\begin{eqnarray}
\nonumber T^{*} &=&
\frac{c\delta}{(1-\eta)\beta(1-\varepsilon)p},\;\;
I^{*}=\frac{(1-\varepsilon)(1-\eta)\lambda p
\beta-dc\delta}{(1-\eta)(1-\varepsilon)p\beta\delta}  \\
        V^{*}    &=& \frac{(1-\eta)(1-\varepsilon)\lambda p
\beta-dc\delta}{(1-\eta)\delta c \beta}=
\frac{p(1-\varepsilon)}{c}I^{*} . \label{infected}
\end{eqnarray}

\indent
 The basic reproduction number $R_{0}$ has been defined in
the introduction as the average number of secondary infections that
occur when one infective is introduced into a completely susceptible
host population \cite{Diek, Dietz, vander}. Note that $R_{0}$ is
also called the basic reproduction ratio \cite{Diek} or basic
reproductive rate \cite{12}. It is implicitly assumed that the
infected outsider is in the host population for the entire
infectious period and mixes with the host population in exactly the
same way that a population native would mix. Following the method
done by \cite{vander}, we have :
\begin{proposition}
The basic reproduction number $R_{0}$ for model (\ref{dyna:gp}) is
given by :
$$ R_{0} = (1-\eta)(1-\varepsilon) \frac{\lambda p \beta}{ c d \delta}   .$$
\end{proposition}
\indent Now we can express the components of infected equilibrium in
term of $R_{0}$. Hence (\ref{infected}) becomes :
\begin{equation}\label{infectedR0}
 T^{*} =\frac{1}{R_{0}}\frac{\lambda}{d},\;\;
I^{*}=\frac{cd}{(1-\eta)(1-\varepsilon)p\beta}(R_{0}-1),\;\;
 V^{*}= \frac{d}{(1-\eta)(R_{0}-1) \beta}
\end{equation}
 \indent
 The following results summarize the main results regarding
the local stability of the disease-free steady state $E^{0}$,
 and
the local stability of the infected steady state during therapy
$E^{*}$.
 The proof of these results can be found in \cite{chong}.
\begin{theorem}
The infection-free steady state $E^{0}$ of model (\ref{dyna:gp})
 is locally asymptotically stable if $R_{0} \leq 1$ and unstable if
$R_{0} > 1$
\end{theorem}
\begin{theorem}
The infected steady state during the therapy $E^{*}$ of model
(\ref{dyna:gp})
 is locally asymptotically stable if $R_{0} > 1$ and unstable if
$R_{0} > 1$
\end{theorem}
\subsection{Global Stability}
In this section, firstly we prove the
global stability of the infection-free equilibrium $E^{0}$ of model
(\ref{dyna:gp}) when the basic reproduction number is less than or
equal to unity. And secondly we prove the global stability of
infected equilibrium $E^{*}$ whenever it exists. We have seen
previously\cite{chong} that the unique positive endemic equilibrium
exits when the basic reproduction number is greater than or equal to
unity.
\begin{theorem}
\begin{enumerate}
    \item[(i)]The infection-free steady state $E^{0}$ of model
    (\ref{dyna:gp}) is globally
    asymptotically stable if the basic reproduction number $R_{0} \leq
    1$ and unstable if $R_{0}> 1$.
    \item[(ii)]The infected steady state during therapy $E^{*}$
     of model
    (\ref{dyna:gp}) is globally
    asymptotically stable if the basic reproduction number $R_{0} \leq
    1$.
\end{enumerate}
\end{theorem}
\begin{prev}
\begin{enumerate}
    \item[(i)]Consider the Lyapunov function :
    $$ L_{1}(T,I,V)=T-T^{0}-
    T^{0}\ln\frac{T}{T^{0}} + I+
     \frac{\delta}{(1-\varepsilon)p}V   .$$
$L_{1}$ is defined, continuous and positive definite for all $T>0$,
$I>0$, $V>0$. Also, the global minimum $L_{1}=0$ occurs at the
infection free equilibrium $E^{0}$. Further,  function $L_{1}$,
along the solutions of system (1), satisfies :
\begin{equation*}
\frac{dL_{1}}{dt} = \lambda-dT-\frac{T^{0}}{T} \lambda + dT^{0}+
  (1-\eta)\beta V T^{0}- \frac{c \delta}{(1-\varepsilon)p}V.
\end{equation*}
Further collecting terms, we have
\begin{equation*}
\frac{dL_{1}}{dt} =\lambda
\left(2-\frac{T}{T^{0}}-\frac{T^{0}}{T}\right)+(R_{0}-1)\frac{c
\delta}{(1-\varepsilon)p}V.
\end{equation*}
Since the arithmetical mean is greater than or equal to the
geometrical mean,$$ \sqrt{a_{1}a_{2}}-(a_{1}+a_{2})\leq 0, \;
a_{i}\geq 0, \; i=1,2.
$$ the function $2-\frac{T}{T^{0}}-\frac{T^{0}}{T}$ are non-positive
for all $T> 0$. In addition, since $R_{0}\leq 1$ ensures
$\frac{dL_{1}}{dt}\leq 0$ for all $T>0$, $V>0$. The equality
$\frac{dL_{1}}{dt}= 0$ holds only (a) at the free equilibrium
$E^{0}$ or (b) when $R_{0}=1$ and $T=T^{0}$. The latter case implies
$I=V=0$.\\
Therefore, the largest compact invariant subset of the set
$$  M= \{ (T, I, V)\in \Omega : \frac{dL_{1}}{dt}= 0 \} $$
is the singleton $\{E^{0}\}$. By the Lasalle invariance
principle\cite{khalil}, the infection-free equilibrium is globally
asymptotically stable if $R_{0}\leq 1$. We have seen previously that
if $R_{0} > 1$, at least one of the eigenvalues of the Jacobian
matrix evaluated at $E^{0}$ has a positive real part.
Therefore, the
infection-free equilibrium $E^{0}$ is unstable when $R_{0}>1$.
\end{enumerate}
\begin{enumerate}
    \item[(ii)]Consider the Lyapunov function :
    \begin{equation*}
L_{2}(T,I,V) = T-T^{*}-T^{*}\ln\frac{T}{T^{*}}+
I-I^{*}-I^{*}\ln\frac{I}{I^{*}}
+\frac{\delta}{(1-\varepsilon)p}\left(V-V^{*}-V^{*}\ln\frac{V}{V^{*}}\right).
    \end{equation*}
The time derivative of $L_{2}$ along the trajectories
of system
(\ref{dyna:gp}) is :
\begin{eqnarray*}
  \frac{d L_{2}}{dt}  &=&
  \frac{dT}{dt}-\frac{T^{*}}{T}\frac{dT}{dt}+
  \frac{dI}{dt}-\frac{I^{*}}{I}\frac{dI}{dt}+\frac{\delta}{(1-\varepsilon)p}\frac{dV}{dt}
  -\frac{\delta}{(1-\varepsilon)p}\frac{V^{*}}{V}\frac{dV}{dt}. \\
\end{eqnarray*}
Collecting terms, and canceling identical terms with opposite signs,
yields
\begin{eqnarray*}
  \frac{d L_{2}}{dt}
    &=& \lambda-dT-\lambda\frac{T^{*}}{T}+dT^{*}+(1-\eta)\beta V T^{*}-(1-\eta)\beta VT \frac{I^{*}}{I}+\delta I^{*} \\
    & &-\frac{cV\delta}{(1-\varepsilon)p}-\delta \frac{V^{*}}{V}I+\frac{c\delta}{(1-\varepsilon)p}V^{*}  \\
    &=& dR_{0}T^{*}-dT^{*}\frac{T}{T^{*}}-dR_{0}T^{*}\frac{T^{*}}{T}+dT^{*}+(1-\eta)\beta V^{*} T^{*}\frac{V}{V^{*}}+\delta I^{*} \\
    & &-(1-\eta)\beta
    V^{*}T^{*}\frac{T}{T^{*}}\frac{V}{V^{*}}\frac{I^{*}}{I}-\frac{c\delta}{(1-\varepsilon)p}\frac{V}{V^{*}}V^{*}
    -\delta \frac{V^{*}}{V}\frac{I}{I^{*}}I^{*}+\frac{c\delta}{(1-\varepsilon)p}V^{*},  \\
    & & \mbox{according to (\ref{infectedR0})},  \\
    &=& dT^{*}\left(R_{0}-\frac{T}{T^{*}}-R_{0}\frac{T^{*}}{T}+1\right) + (1-\eta)\beta V^{*}T^{*}\left(\frac{V}{V^{*}}
    -\frac{VTI^{*}}{V^{*}T^{*}I}\right)\\
    & & +\delta I^{*}\left(2+\frac{V}{V^{*}}-\frac{V^{*}I}{VI^{*}}\right) \\
    &=&  dT^{*}\left(R_{0}-\frac{T}{T^{*}}-R_{0}\frac{T^{*}}{T}+1\right) + \delta I^{*}\left(\frac{V}{V^{*}}
    -\frac{VTI^{*}}{V^{*}T^{*}I}\right)\\
    & &+\delta I^{*}\left(2+\frac{V}{V^{*}}-\frac{V^{*}I}{VI^{*}}\right), \; \mbox{since $(1-\eta)\beta V^{*}T^{*}= \delta I^{*}
    $}
\end{eqnarray*}
\begin{eqnarray*}
  \frac{d L_{2}}{dt}
    &=& dT^{*}\left(R_{0}-\frac{T}{T^{*}}-R_{0}\frac{T^{*}}{T}+1\right) + \delta I^{*}\left(2-\frac{V^{*}I}{VI^{*}}
    -\frac{VTI^{*}}{V^{*}T^{*}I}\right) +\delta I^{*}\left( 1-\frac{T^{*}}{T}\right)\\
    && - \delta I^{*}\left( 1-\frac{T^{*}}{T}\right)\\
    &=& dT^{*}\left(R_{0}-\frac{T}{T^{*}}-R_{0}\frac{T^{*}}{T}+1\right) + \delta I^{*}\left(3-\frac{T^{*}}{T}-\frac{V^{*}I}{VI^{*}}
    -\frac{VTI^{*}}{V^{*}T^{*}I}\right) \\
    && -d T^{*}(R_{0}-1)\left( 1-\frac{T^{*}}{T}\right),\; \mbox{since $\delta
    I^{*}=dT^{*}(R_{0}-1)$},\\
\frac{d L_{2}}{dt} &=&
dT^{*}\left(2-\frac{T}{T^{*}}-\frac{T^{*}}{T}\right) + \delta
I^{*}\left(3-\frac{T^{*}}{T}-\frac{V^{*}I}{VI^{*}}
    -\frac{VTI^{*}}{V^{*}T^{*}I}\right).\\
   & \leq & 0.
\end{eqnarray*}
The terms between the brackets are less than or equal to zero by the
inequality (the geometric mean is less than or equal to the
arithmetic mean)
$$ \sqrt[3]{a_{1}a_{2}a_{3}}-(a_{1}+a_{2}+a_{3}) \leq 0,
 \;\; a_{i}\geq 0,\; i=1, 2, 3     .$$
It should be noted that $\frac{d L_{2}}{dt} = 0 $
holds if and only
if $(T, X, V)$ take the steady states values
$(T^{*}, X^{*}, V^{*})$
Therefore the infected equilibrium $E^{*}$ is globally
asymptotically stable.
\end{enumerate}
\end{prev}
\section{Numerical Simulation}
Some numerical simulations have been done in the case $ R_{0}< 1 $
to confirm theoretical result obtain on global stability for the
uninfected equilibrium.

\begin{figure}[!h]
\centering
\begin{subfigure}[b]{0.3\textwidth}
\includegraphics[angle=0,height=5cm,width=\textwidth]
{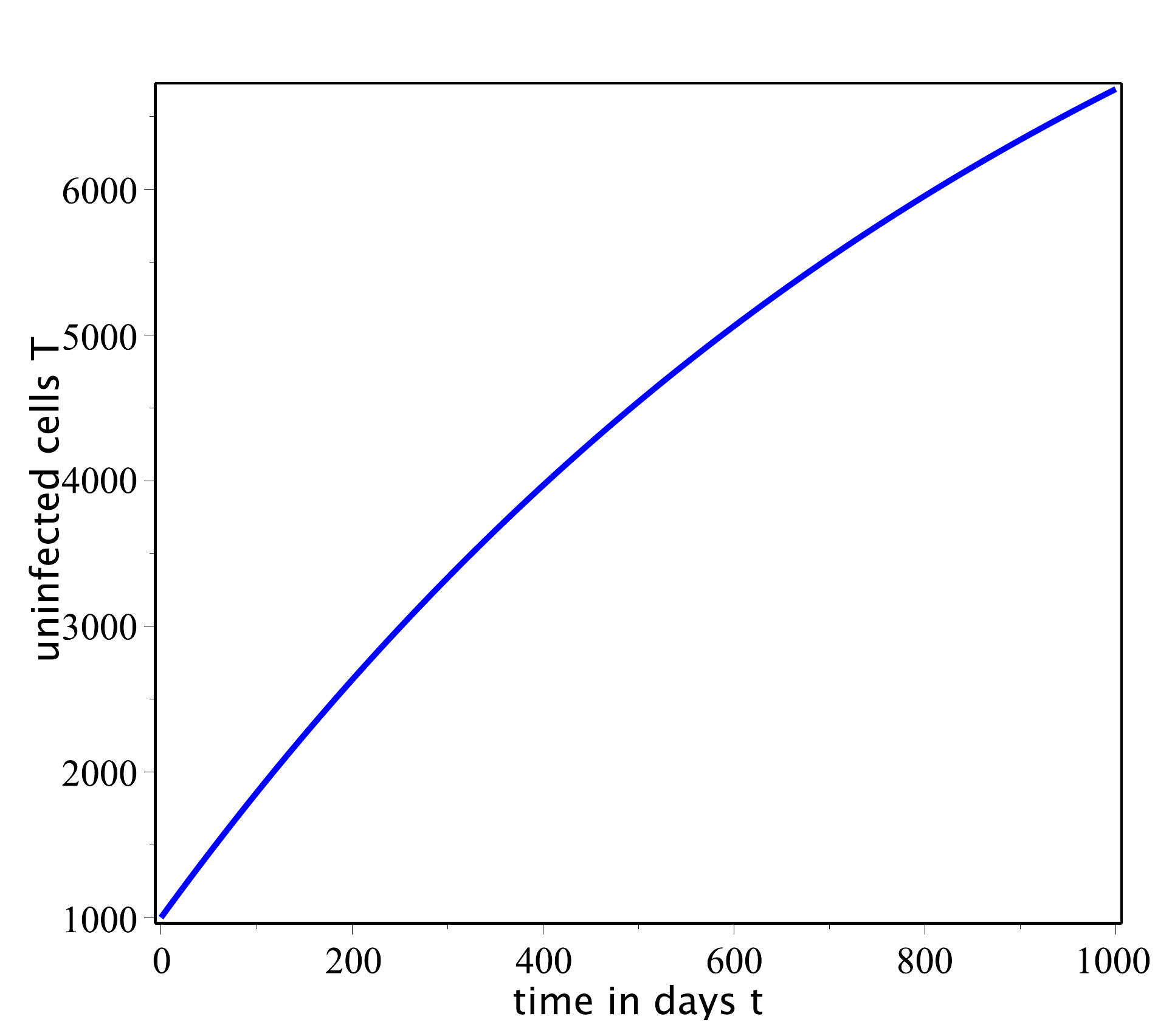} \subcaption{solution curve for the uninfected
cells}
\end{subfigure}
\begin{subfigure}[b]{0.3\textwidth}
\includegraphics[angle=0,height=5cm,width=\textwidth]{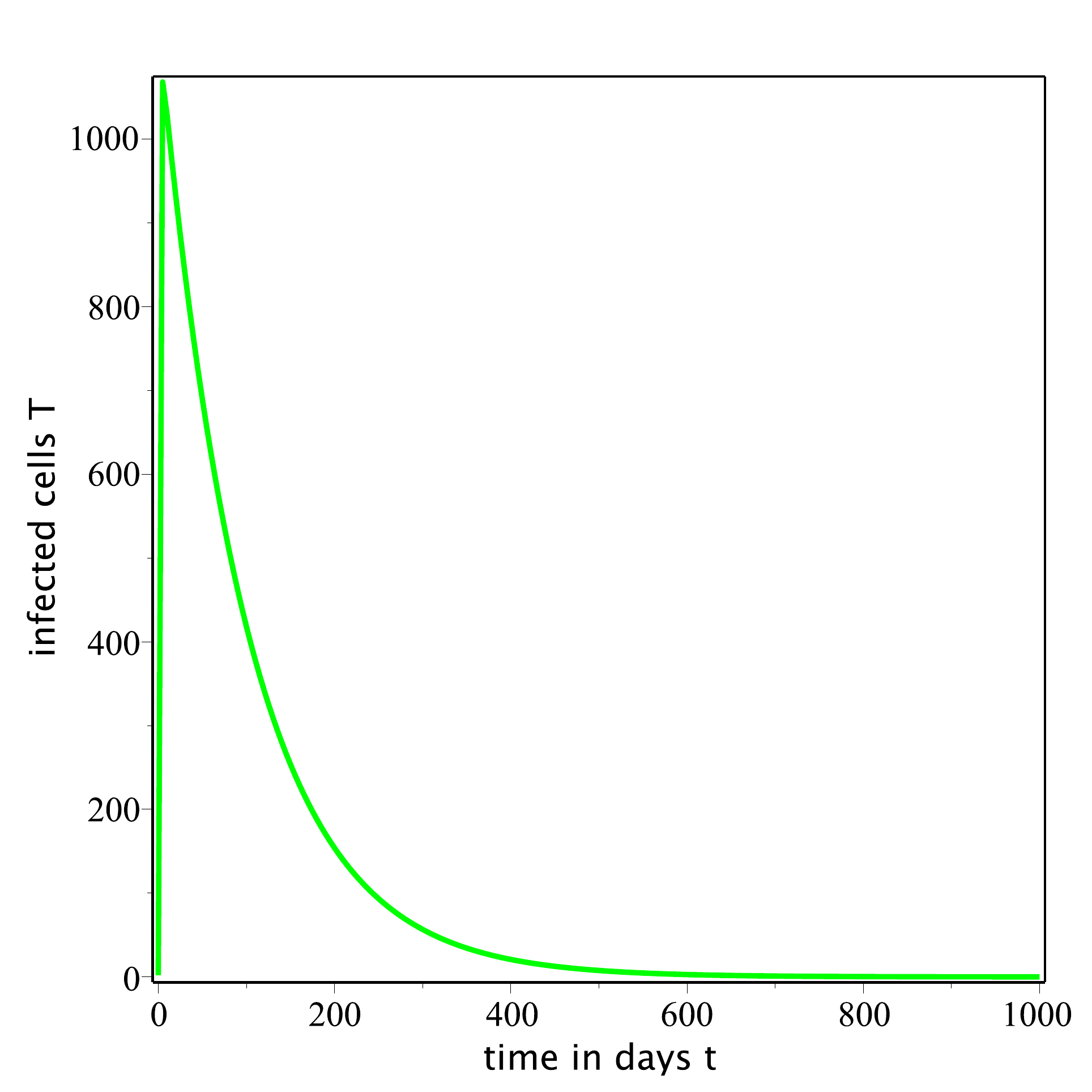}
\subcaption{numerical solution curve for the infected cells}
\end{subfigure}
\begin{subfigure}[b]{0.3\textwidth}
\includegraphics[angle=0,height=5cm,width=\textwidth]{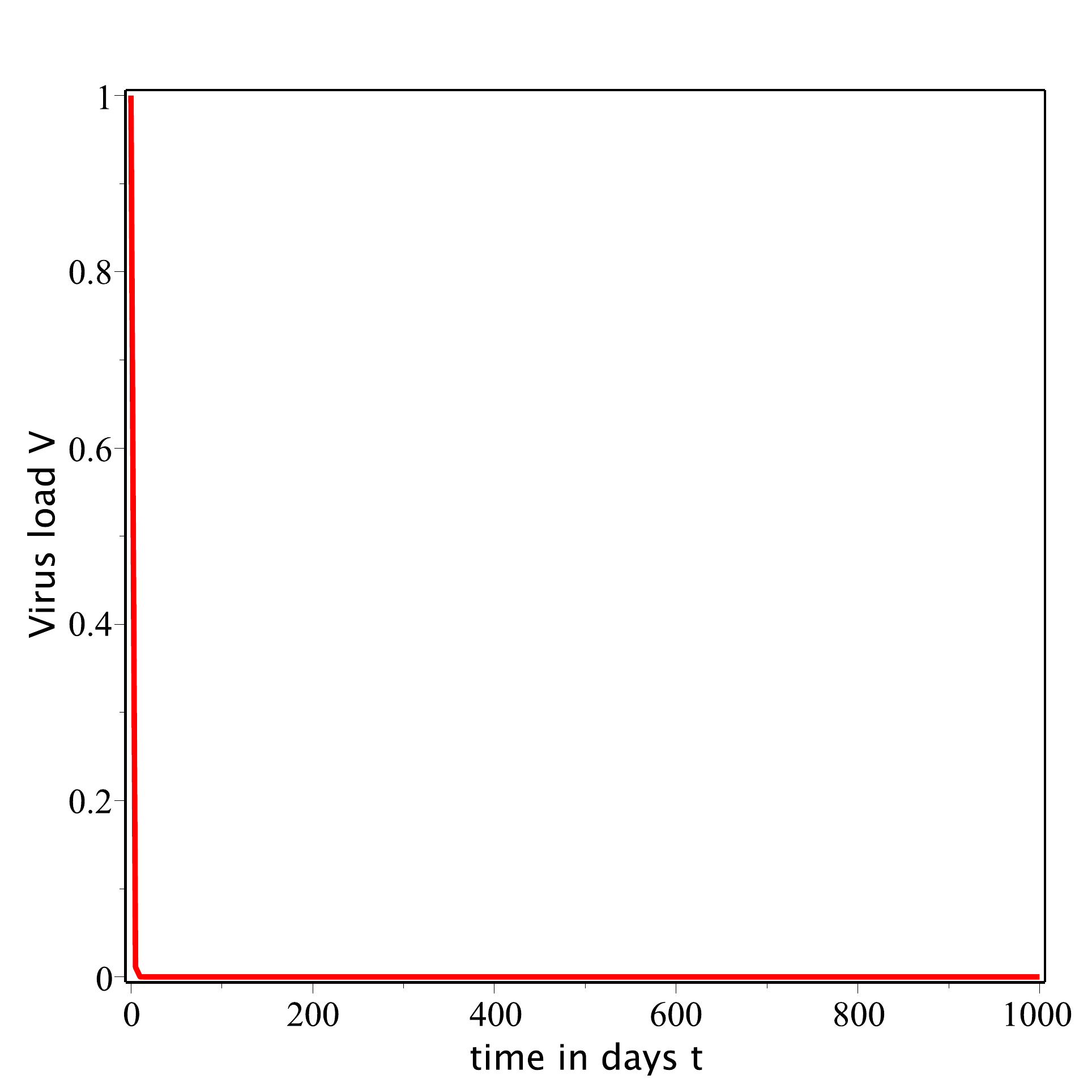}
\subcaption{numerical solution curve for the virus load}
\end{subfigure}
\end{figure}
\begin{figure}[!h]
\centering
\includegraphics[angle=0,height=5cm,width=10cm]{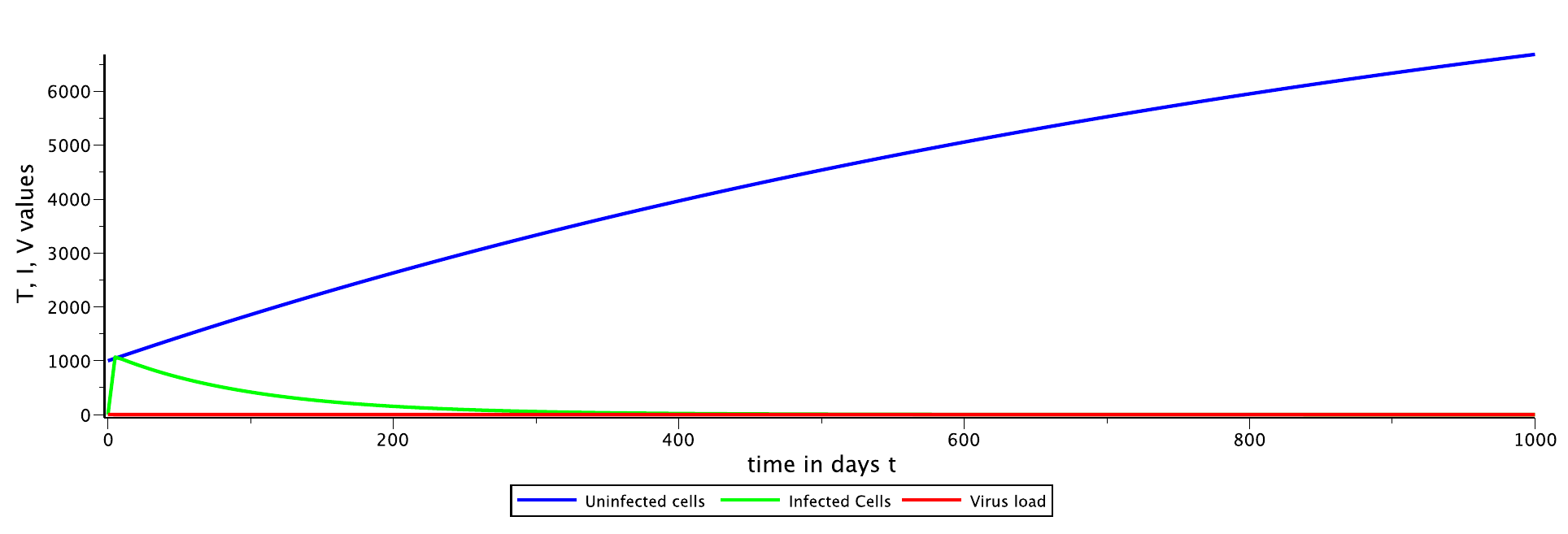}
\caption{numerical simulation of the HCV model in 1000 days}
\end{figure}
 We run simulations using the initial conditions:
 $T=10^{3}$, $I=2$ and $V=1$ and the following parameter
 values : $\lambda = 10 $; $ \eta = \frac{1}{10000000} $;
 $\beta = 2.4 \times 10^{-8} $; $\delta = 0.01$;
 $ d= 0.001$ ; $\varepsilon = 0.00000001 $; $c= 0.9$,
  $p= 0.000000001 $.
\newpage
\section{Conclusion}
In this paper, we have extended the first part of the work done by
Chong et al. in \cite{chong} where they only studied the local
stability of the fundamental mathematical model of hepatitis C
infection with treatment.
\section*{Acknowledgements}
I am  grateful to Professor Alan Rendall for valuable and tremendous
discussions. I wish to thank him for introducing me to Mathematical
Biology and to its relationship with Mathematical Analysis.

\end{document}